\documentclass[a4paper, 12pt]{article}
\usepackage[english]{babel}
\usepackage[a4paper, inner=2cm, outer=2cm, top=3cm, bottom=3cm]{geometry}
\usepackage{mathtools}
\usepackage{pstricks}
\usepackage{caption}
\usepackage{graphicx}
\usepackage{amsmath}
\usepackage{amssymb}
\usepackage{mathcomp}
\usepackage{textcomp}
\usepackage{physics}
\usepackage[doublespacing]{setspace}

\title{On the cutoff identification and the quantum improvement in asymptotically safe gravity}
\author{R. Moti and A. Shojai\\
\textit{\small Department of Physics, University of Tehran, Tehran, Iran}}
\date{}
\begin{document}
\maketitle
\pagenumbering{arabic}
\begin{abstract}
Applying the exact renormalization group method to search the non--Gaussian fixed points of gravitational coupling, is frequently followed by two steps: \textit{cutoff identification} and \textit{improvement}. Although there are various models for identifying the cutoff momentum by some physical length, saving the general covariance should be considered as an important property in the procedure. 
In this paper, use of an arbitrary function of curvature invariants for cutoff identification is suggested. It is shown that the field equations for this approach differs from the ones obtained from the conventional cutoff identification and improvement, even for non--vacuum solutions of the improved Einstein equations. Indeed, it is concluded that these two steps are correlated to each other.
\end{abstract}

\section{Introduction}
Finding a consistent quantum gravity theory which fulfills the cornerstones of general relativity and quantum mechanics, simultaneously, is still a big challenge for theoretical physicists. A coordinate independent theory which could describe the dynamical quanta of gravitational field does not proposed yet.

The canonical quantization of gravitational interaction suffers from divergences which can not be renormalized by the known renormalization methods, using a finite number of counter terms. Although the compatibility of the usual quantum field theory with general relativity is still a debatable issue \cite{Rovelli}, but the suggestion of a new approach to eliminate the UV divergences of the quantized general relativity theory could be useful, even for other quantum field theories.

In this sense, the Weinberg's \textit{asymptotic safety conjecture} \cite{Weinberg-1st} is considerable since the asymptotically free theories can be categorized as a kind of asymptotic safe ones.
By this conjecture, the theory would be renormalizable  and a predictive one provided that it has a non--Gaussian UV fixed point and all the essential running couplings' trajectories lie on a finite dimensional critical surface of this fixed point and tend to it at the UV limit \cite{finite-dimensional} (see references in \cite{Weinberg inflation}).

Using the exact renormalization group equation (ERGE) for the gravitational flow, $\Gamma_k$ \cite{Reuter-1st}, a running gravitational coupling, which satisfies the asymptotic safety conjecture can be found \cite{Souma}.

Studying the effects of this renormalization method on the low energy scales, needs the solutions of ERGE. But this process is complex and maybe practically impossible because of the multiplicity of flow terms. Hence, one usually applies the truncation method and project the flow onto a given sub--theory space and extract the $\beta$--functions. To apply the theory to physical phenomena, fully quantum effective action is needed, which is not available. Hence, this functional renormalization group method is usually followed by two steps: cutoff identification and improvement.

By cutoff identification one means that, since the gravitational interaction describes the space--time, the renormalization group scaling parameter $k$ should be identified by a function of space--time distances such as the inverse of physical length which could define the space--time scale.
Then, improving the coupling constant of the classical theory to the running one, the quantum corrections of this method at the low energy could be studied.
Since these coupled steps could seriously affect the general covariance of the action and thus the background independence of the theory, they need  more studies.

There are many suggestions to search a possible non--Gaussian UV fixed point for gravitational running coupling which saves the theory from UV divergences (\cite{finite-dimensional, Reuter-1st, Souma, ERGE} and see references in \cite{Weinberg inflation}).

Introducing the gravitational renormalization group flow, Reuter suggests an anti--screening behaviour for gravitational coupling near non--Gaussian fixed point \cite{Reuter-1st}. This method, uses the exact renormalization group equation (ERGE) of Wetterich \cite{Wetterich},
 
\begin{equation}
k\partial_k \Gamma_k = \frac{1}{2} \Tr\bigl[ \bigl(\Gamma^{(2)}_k+\mathcal{R}_k\bigr)^{-1} k\partial_k\mathcal{R}_k \bigr] \ ,
\end{equation}
to find the running couplings. This equation describes the evolution of the effective average action (EAA), $\Gamma_k$, by the passage of the renormalization group time, $t \equiv k\ln k$. 

This scale dependent effective action, $\Gamma_k \equiv \Gamma_k[g_{\mu\nu}]$ near a mass scale $k$ is obtained from the bare action by integrating out all quantum fluctuations with momenta larger than the infrared cutoff $k$ \cite{Reuter & Weyer 1}. The $\Gamma^{(2)}_k$ is the Hessian of $\Gamma_k$ and the $\mathcal{R}_k$ is the IR--cutoff.
The arbitrary smooth function  $\mathcal{R}_k$ suppresses the low momentum modes, however, it is required that $\mathcal{R}_{k\rightarrow\infty}$ vanishes in  order to not disturb high momentum modes. The Ward identity requires that $\mathcal{R}_{k}$ be a quadratic function of momentum $k$ \cite{Reuter-1st}. Hence, if  $\mathcal{R}_k(p^2) \propto k^2 \mathcal{R}^{(0)}(p^2/k^2) $, the function $ \mathcal{R}^{(0)}_k (\kappa) $ should satisfies the conditions $ \mathcal{R}^{(0)}(0) =1 $ and $ \mathcal{R}^{(0)}(\kappa \rightarrow \infty) \rightarrow 0$. The exponential form $ \mathcal{R}^{(0)}(\kappa) = \kappa/(\exp\kappa -1) $ might be a proper chosen form which is used in the literature \cite{Nagy}.

The EAA at high momentum scales is considered as an initial condition for this differential equation. This initial condition evolves trough the ERGE at the IR extreme. From the definition of EAA it is taken that $\Gamma_{k\to\infty}\equiv S$ and $\Gamma_{k\to 0}\equiv \Gamma$ where $\Gamma$ is the classical effective action. The classical action $S$ which is considered as an initial condition is the one with the coupling constant improved to running unspecified one. Applying this initial condition to ERGE, ends to multi $\beta$--functions which are running couplings' evolution equations. 

As a result of truncation, we encounter a system of solvable partial differential equations. For example, to find the running gravitational coupling, the Einstein--Hilbert truncation,
\begin{equation}
\Gamma_k = \frac{1}{16\pi G(k)} \int\dd[4]{x} \sqrt{g} \bigl(-R+2\Lambda(k) \bigr)
\end{equation}
seems suitable for searching for the effects of this renormalization method on the metric. The numerical solution of this $\beta$--function at \textit{infrared regime} ($k\rightarrow0$), where the elimination of quantum fluctuation modes recovers the perturbative RG treatment, can be found to be $  G(k) = G_0 \left[ 1 - \omega G_0 k^2 + \mathcal{O}(G_0^2 k^4)\right]$, while near the \textit{fixed point} ($k\gg m_{pl}$), it is $G(k) = g_{*}^{UV}/{k^2}$. These two numerical solutions can be combined and written as
\begin{equation}
  G(k) = \frac{G(k_{0})}{1 + \omega G(k_{0})(k^{2}-k_{0}^{2})}
\end{equation}
where $ k_{0} $ is a reference scale, with the condition $ G_{N} \equiv G(k_{0} \to 0)=G_0$
in which  $G_N$ is the experimentally observed value of Newton's constant, and $\omega = \frac{4}{\pi}(1-\frac{\pi^{2}}{144})$ \cite{Bonanno & Reuter 1, Bonanno & Reuter 3, Bonanno & Reuter 2}.

It is clear that the effect of low--momentum terms such as non--local terms are eliminated by in this process. Hence the general $\Gamma_{k\to 0}$ cannot be accessed to study the quantum metric $g_{\alpha\beta} \equiv \expval{g_{\alpha\beta}}$ from
\begin{equation}
\eval{\fdv{\Gamma[g_{\alpha\beta}]}{g_{\alpha\beta}}}_{g_{\alpha\beta}\equiv\expval{g_{\alpha\beta}}}=0 \ .
\end{equation}

It is claimed that improvement of typical coupling constant $g_0$ to running one $g_k$ in the classical theory, can restore the effects of quantum loops to some extent \cite{Reuter & Weyer 1}.

In what follows, after a brief introduction of a cutoff identification, a new identification is suggested, which make the action improvement possible without losing the general covariance. It is argued that the cutoff identification should be in general a function of all the 20 curvature invariants. Although for some specific identifications (like $R, R_{\alpha\beta}R^{\alpha\beta}, R_{\alpha\beta\gamma\delta}R^{\alpha\beta\gamma\delta}, \ldots$) \cite{f(R), Bonanno, Domazet, Moti & Shojai 2, Pawlowski} and for some specific solutions, this is studied, here we make a general discussion of the method. Indeed, the improved Friedman equation for a flat cosmological model filled with cosmological constant is studied and the last section is dedicated to a conclusion on this identification method.

\section{Cutoff identification and improvement} \label{3}
For non--gravitational (e.g. electromagnetic) interactions the IR--cutoff parameter $k$ which is the scaling parameter of the renormalization group, straightly is related to some parameter of either the bound state (e.g. hydrogen atom radius) \cite{Reuter & Weyer 1} or the scattering process (e.g. the impact parameter). Hence, improving $g_0$ to $g_k$ (i.e. $g(k)$) is an acceptable suggestion for considering the quantum fluctuations as quantum corrections to the classical solutions.
But for the gravitational interaction, since the gauge field is the spacetime itself, the cutoff $k$ can not be identified in such a simple way. The acceptable parameter for a gravitational system should be, in some sense, a standard scale by which the comparison between spacetime measurements is done.

In the flat spacetime, the length can be interpreted as a standard scale of comparison and the cutoff identification $k=\xi/x$ seems suitable for this purpose. This is because the Fourier modes in flat spacetime are tantamount to momentum, hence related to the length. But the equivalence of Fourier modes with momentum is not necessarily possible in curved spacetimes. In other words, the identification $k=\xi/x$ cannot be correct for a \textit{general} curved spacetimes. Although for the flat background coordinate dependency do not threaten the covariance of the theory, but for a \textit{general} curved spacetime with such  an identification neither general covariance is gauranteed to be survived nor the scaling role of the identification.

In some cases, the symmetries of metric would be helpful. For example, for cosmological models such as Friedmann--Robertson--Walker(FRW) the identification $k=k(t,a(t),\dot{a}(t), \ldots)$ seems to be enough. Indeed, for Schwarzchild static spherical vacuum solution , the  physical length $d(x)=\sqrt{g_{ij}dx^idx^j}$ can be used to define $k=\xi/d(x)$ as a cutoff identification \cite{Reuter & Weyer 1}. But having a general definition for relating cutoff momentum to an appropriate \textit{scaling} parameter of the general system is necessary.

As mentioned, the cutoff $k$ in EAA should be identified with a scale parameter as the standard for determination of smallness. In general relativity, the physical length is observer dependent and thus is not suitable for this purpose, hence, an observer independent quantity is needed. 
Since the correct explanation of size can be found in the gauge character of gravity, the appropriate quantity with the meaning of smallness should be searched gauge theory of gravity and tidal forces.

The gravitational gauge field, i.e. the affine connection, is the cause of the rotation of local tetrad coordinate basis with respect to the  neighbouring basis. The magnitude of this rotation which determines the gravitational field strength is represented by the Riemann tensor and has the physical meaning of tidal forces experienced by a falling object. Considering two free falling particles separated by the distance $\zeta^{\alpha}$, the tidal forces act according to the geodesic deviation equation
\begin{equation}
\frac{D^2\zeta^{\alpha}}{D\tau^2} = R^{\alpha}_{\ \beta\gamma\delta} \dv{x^{\beta}}{\tau}\dv{x^{\gamma}}{\tau}\zeta^{\delta} \ ,
\end{equation}
where $ R^{\alpha}_{\ \beta\gamma\delta} $ is Riemann curvature tensor and the RHS is the gravitational tidal force. 

It is the tidal force that is the agent of change of distances between particles and provides the absolute meaning of distance, in the sense that small distance is defined as what for it the geodesic deviation is ignorable. As a result \textit{an appropriate function of the tidal forces is a good candidate for cutoff identification}.

There are twenty independent components of $R^{\alpha}_{\ \beta\gamma\delta}$ in four dimensional spacetime, and instead of these tensor components, one can use twenty curvature invariants. There can be various set of curvature invariants depending on what Petrov set is chosen. The most known of them are $ R $, $ R^{\alpha\beta}R_{\alpha\beta}$ and $R^{\alpha\beta\gamma\delta}R_{\alpha\beta\gamma\delta}$ \cite{Petrov}.

By these considerations, we \textit{propose} the following cutoff identification
\begin{equation}
k = \frac{\xi}{\chi(\chi_1,\ldots,\chi_{20})}
\end{equation}
which seems to be the best choice for gravitational systems. $\xi$ is some dimensionless constant and the function $\chi(\chi_1,\ldots,\chi_{20})$ determines the maximum neighbourhood \textit{size} which can be considered so small that the equivalence principle is applicable and no tidal force is observable.
According to the equivalence principle, any gravitational field \textit{locally} can be removed by going to an appropriate frame, the freely falling frame. By \textit{locally} one means in the neighbourhood of any point. For a neighbourhood small enough the tidal forces do not deviate the Euclidean parallel geodesics and the scope of validity of this principle is determined by the tidal forces which are proportional to the Riemann curvature tensor or in terms of $\chi_1,\ldots,\chi_{20}$. Since in the local frame special relativity holds, such an identification seems desirable.
This provides a cutoff identification compatible with general covariance and the equivalence principle.

After the introduction of this new cutoff identification, we have to improve the coupling constant as mentioned previously. Using the above identification, the improvement is done by changing $G_0$ to $G(k(\chi))$ in the classical theory. The improvement may be done in the solutions of the classical theory or in the Einstein's equations themselves. These are called \textit{solution improvement} and \textit{equation of motion (EOM) improvement} methods respectively. There are two other ways of improvement methods. One is to improve the coupling constants in the action, but considering them as input parameters, which we call \textit{parameter improvement}. The last way is improving the action via the above mentioned cutoff identification. This method which we call \textit{action improvement} and is the most physical acceptable method both saves the general covariance of the improved equations and is compatible the equivalence principle. Also in contrast to other mentioned methods for which the improved couplings act as input variables, here they contribute to the improved equations of motion dynamically.

In what follows we compare these improvement methods using the standard and the above proposed cutoff identification.

\subsection{Solution improvement}
Improvement of gravitation coupling constant $G_0$ to the running coupling $G(x)$ in the solutions is called the solution improvement \cite{Reuter & Weyer 1}. Although this improvement has the correct classical limit, but the dynamics of quantum corrections is forgotten in it. It is preferred to use this method when the backreaction effects of the quantum corrections on the background geometry are negligible. The solution improvement is useless for the situations where the quantum corrections are not small like near the black hole singularity.

\subsection{Equation of motion improvement}
The EOM improvement method is improving the coupling constants in the field equation to the running ones, and then finding the solutions. This method is widely used in the litreature (see \cite{Reuter & Weyer 1, Reuter & Tuiran, Falls}). In the Einsteinian gravity theory the gravitational coupling constant in the equations of motion would be improved to $G(x)$. Although this kind of improvement does not considers any dynamics for the coupling constants, like in the solution improvement, but because of improvement of the equations of motion the backreaction effects of quantum corrections on the geometry is included in some way.

It is interesting to notice that in some cases such as the vacuum solutions, this method leads to the same results as the solution improvement method \cite{Reuter & Weyer 1}. The Schwarzschild and Kerr solutions are studied using this improvement method and physical length as the cutoff identification \cite{Reuter & Tuiran, Bonanno & Reuter 1, Bonanno & Reuter 3}. 

Since the cutoff identification is an important step in studying the effects of quantum improvement, let's search for the solutions of the improved Friedmann equation, using different types of cutoff identification functions.

To do so, consider the FRW metric
\begin{equation}
 \dd{s}^2 = c^2 \dd{t}^2 - a(t)^2 \left(\frac{\dd{r}^2}{1-Kr^2} +r^2 \dd{\theta}^2 + r^2\sin^2\theta \dd{\phi}^2\right) \ .
\end{equation}
The behaviour of scale factor $a(t)$ is determined by the Einstein equations $G_{\alpha\beta}= 8\pi G T_{\alpha\beta}$. 
Using the prefect fluid with the equation of state $p=w\rho$ for the energy--momentum tensor, we get the following improved Friedman equations:
\begin{align}
 & 3\frac{\ddot{a}(t)}{a(t)} = -4\pi G(k) \rho( 1 + 3 w) \label{FRW1}\\
 & \frac{\ddot{a}(t)}{a(t)} + 2\frac{\dot{a}(t)^2}{a(t)^2} +2\frac{c^2 K}{a(t)^2} = 4\pi G(k) \rho (1 - w) \label{FRW2} \ .
\end{align}

\begin{itemize}

\item \textbf{Naive identification}($\mathbf{k=\xi/ct}$):

Amonge all the proposed identifications (such as $k^4\sim \rho$ \cite{Bonanno & Koch}, $k\sim 1/d(r)$ \cite{Bonanno & Reuter 3, Adeifeoba & Astrid}, $k\sim H(t)$ \cite{Bonanno & Reuter 4}, etc.) for cosmological solutions, one can argue that the cosmological time and the scale factor make the identification $k = k(t,a(t),\dot{a}(t),\ldots)$ a proper one. For homogeneous and isotropic cosmos this can be reduced to $k=k(t)$ \cite{Bonanno & Reuter 2}, \cite{Sola}. So, if $\xi$ is a dimensionless constant, the identification $k=\xi/ct$ leads to the improvement
\begin{equation}
 G_0 \rightarrow G(k) = G(\xi/ct) = \frac{G_{0}}{1+ \omega G_{0}\xi^2 /(ct)^2} = \frac{G_{0}}{1+ \mathcal{A}/t^2}
 \end{equation}
with $\mathcal{A} \equiv  \omega G_{0} \xi^2/c^2 $. 
Substituting this in the equations \eqref{FRW1} and \eqref{FRW2} results in
\begin{align}
  & 3\frac{\ddot{a}}{a}( 1+ \mathcal{A}/t^2) =-4\pi G_{0}\rho( 1 +3 w)  \ , \\
  & (\frac{\ddot{a}}{a} + 2\frac{\dot{a}^2}{a^2} +2\frac{c^2 K}{a^2})  ( 1+ \mathcal{A}/t^2)= 4\pi G_{0}\rho( 1 - w) \ .
\end{align}
For flat ($k = 0$) universe filled with cosmological constant ($w_{\Lambda}=-1$), and setting $ a(t)\equiv a_1(t) = a_* e^{\alpha_1(\tau)} $, where $\tau=t/t_*$ we have
\begin{equation}
 \frac{2}{t_*^2}\alpha''_1(1 + \frac{\mathcal{A} t_{*}^{2}}{\tau^2}) = 0 \ .
\end{equation}
The prime indicates derivative with respect to dimensionless time $\tau$. Since $\mathcal{A}>0$, the second factor does not have any real root, the solution is $ \alpha_1(\tau)=C_1\tau+C_2 $ with $C_1$ and $C_2$ constants. Therefore one gets the classical de Sitter scale factor
\begin{equation}
a_1(t) = a_i e^{t/t_i}
\end{equation}
with initial conditions $a_i$ and $t_i$. As expected, the parameter $\mathcal{A}$ which contains the quantum running parameter $\omega$, does not contribute to the dynamics of the scale factor. But this does not mean that the solution is classical, as the quantum effects appear in the matter density as (obtained from the Friedman equations):
\begin{equation}
\rho_1 =\rho_i ( 1 + \mathcal{B}\frac{t_i^2}{t^2} ) \ ,
\end{equation}
where $\rho_i \equiv 8\pi G_0t_i^2/3 $, $\rho_1 \equiv \rho(a_1(t))$ and $ \mathcal{B} \equiv \mathcal{A}/t_i^2 $. 
For times $ t^2\ll \mathcal{B}t_i^2 $, where high energy modes are important, quantum effects on the matter density is considerable, while it is ignorable for $ t^2 \gg \mathcal{B}t_i^2 $.

We have found this solution for a universe filled with cosmological constant, but similar effects can be seen for more realistic cases. For example,
the solution of the improved Friedman equation with standard cutoff identification and non--zero matter density is studied in \cite{Reuter & Weyer 1}. It is shown that the scale factor is proportional to $t^{2/(3+3w)}(1+\mathcal{O}(t^{-2}))$, and the matter density behaves as $t^{-2}(1+\mathcal{O}(t^{-2}))$. In \cite{Moti & Shojai 1}, the effect of this kind of improvement on the cosmological perturbation equation and the resulting power spectrum are studied. It is shown that the quantum corrections are important for high momentum modes and are consistent with observations.
 
\item \textbf{Identification with }$\mathbf{k = \xi R^{1/2}}$:

The identification  $\chi=R^{-1/2}$ is the one which can be used to find a related $f(R)$ theory \cite{f(R)}, let us use it for the cosmological FRW solution \cite{Bonanno, Domazet}.
Since the scalar curvature of FRW metric is $ -6 (a\ddot{a}+\dot{a}^2+Kc^2)/(c^2 a^2) $, the improvement gives
\begin{equation}
G_0 \rightarrow G(k) = G(\xi R^{1/2}) = \frac{G_{0}}{1+ \omega G_{0}\xi^2 R} = \frac{G_{0}}{1- 6\mathcal{A}(\frac{\ddot{a}}{a}+\frac{\dot{a}^2}{a^2}+\frac{Kc^2}{a^2} )} \ .
\end{equation}
Substituting this in the equations \eqref{FRW1} and \eqref{FRW2} leads to
\begin{align}
  & 3\frac{\ddot{a}}{a} (1- 6\mathcal{A}(\frac{\ddot{a}}{a}+\frac{\dot{a}^2}{a^2}+\frac{Kc^2}{a^2} )) =-4\pi G_{0}\rho( 1 +3 w) \ ,\\
  & (\frac{\ddot{a}}{a} + 2\frac{\dot{a}^2}{a^2} +2\frac{c^2 K}{a^2})(1- 6\mathcal{A}(\frac{\ddot{a}}{a}+\frac{\dot{a}^2}{a^2}+\frac{Kc^2}{a^2} ))= 4\pi G_{0}\rho( 1 - w) \ .
 \end{align} 
In a similar way to the previous case, for flat universe filled with the cosmological constant, the solution is $ a(t) \equiv a_2(t) = a_* e^{\alpha_2(\tau)} $, where
\begin{equation} \label{FRW-2ndCI}
 \alpha''_2(1 -6 \frac{\mathcal{A}}{ t_*^2}(\alpha''_2+2\alpha'^2_2)) = 0 \ .
\end{equation}
There are two solutions. First setting $\alpha_2''=0$, we get de Sitter solution with a constant matter density
\begin{equation}
\rho_2 = \rho_i (1-12 \mathcal{B})  \ .
\end{equation}
This is just the classical solution.
Since $\rho_2$ is assumed to be positive, we should have $\mathcal{B}>1/12$.

In addition to the classical de Sitter solution $a_2(t)$, there is another solution given by setting
\begin{equation}
 \alpha''_2 + 2 \alpha'^2_2 = \frac{t_*^2}{6\mathcal{A}} \ ,
\end{equation}
in the equation \eqref{FRW-2ndCI}. The scale factor is then
\begin{equation}
 a^{(\mathcal{A})}_2(t) = a_i \cosh^{1/2}(\frac{t}{\sqrt{3\mathcal{A}}}+\tau_i) \ ,
\end{equation}
with initial conditions $\tau_i $ and $a_i$. Surprisingly, the matter density $\rho_2^{(\mathcal{A})}$ for this solution vanishes for all possible initial conditions. Clearly this solution is a pure quantum solution. 

Rewriting $a^{(\mathcal{A})}_2(t)$ as $ a_i(C e^{t/\sqrt{3\mathcal{A}}}+ e^{-t/\sqrt{3\mathcal{A}}} )$, for $C=0$ the anti--de Sitter scale factor is obtained. The tendency to this solution comes from anti--screening behaviour of the running gravitational coupling $G(k)$. 
From anti--screening property we expect that at low energy the quantum effects of gravity overcome the accelerated expansion of vacuum which is fulfilled by anti--de Sitter solution.

\item \textbf{Identification with} $\mathbf{k = \xi (R_{\alpha\beta}R^{\alpha\beta})^{1/4}}$:

As another example, let us choose $\chi = (R_{\alpha\beta}R^{\alpha\beta})^{-1/4}$, which has been studied for Schwarzschild vacuum solution \cite{Moti & Shojai 2}. This identification ends to
\begin{align}
 G_0 \rightarrow G(k) = G(\xi (R_{\alpha\beta}R^{\alpha\beta})^{1/4}) & = \frac{G_0}{1+ \omega G_{0}\xi^2 (R_{\alpha\beta}R^{\alpha\beta})^{1/2}}\nonumber \\ 
 &= \frac{G_{0}}{1+ \mathcal{A}(9 a^2 \ddot{a}^2 + 3 (a\ddot{a}+2\dot{a}^2+2Kc^2)^{2})^{1/2}/a^2}
\end{align}
for FRW with $ R_{\alpha\beta}R^{\alpha\beta} = (9 a^2 \ddot{a}^2 + 3 (a\ddot{a}+2\dot{a}^2+2 Kc^2)^{2})/(c^4a^4) $. 

Substituting this $G(k)$ in the equations \eqref{FRW1} and \eqref{FRW2} gives the modified Friedman equations:
\begin{align}
  & 3\frac{\ddot{a}}{a} (1+ \mathcal{A}\frac{(9 a^2 \ddot{a}^2 + 3 (a\ddot{a}+2\dot{a}^2+2Kc^2)^{2})^{1/2}}{a^2} ) =-4\pi G_{0}\rho( 1 + 3 w) \ , \\
  & (\frac{\ddot{a}}{a} + 2\frac{\dot{a}^2}{a^2} +2\frac{c^2 K}{a^2})(1+ \mathcal{A}\frac{(9 a^2 \ddot{a}^2 + 3 (a\ddot{a}+2\dot{a}^2+2Kc^2)^{2})^{1/2}}{a^2} ) = 4\pi G_{0}\rho( 1 - w) \ . 
\end{align}
Again for flat universe filled with cosmological constant, the solution can be obtained to be $ a(t) \equiv a_3(t) = a_0 e^{\alpha_3(\tau)} $, where
\begin{equation}
 \alpha_3''(1 + 2\sqrt{3} \frac{\mathcal{A}}{ t_{0}^{2}}(\alpha_3''^{2}+3\alpha_3''\alpha_3'^{2}+3\alpha_3'^{4})^{1/2}) = 0 \ .
\end{equation}
This differential equation has only a de Sitter solution, as it was the case for the standard identification. The matter density equals to
\begin{equation}
\rho_3 = \rho_i (1+6 \mathcal{B})
\end{equation}
where $ \mathcal{B} \equiv  \mathcal{A}t_i^2 $.

\item \textbf{Identification with} $\mathbf{k = \xi (R_{\alpha\beta\gamma\delta}R^{\alpha\beta\gamma\delta})^{1/4}}$:

Finally, we consider the identification $\chi=(R_{\alpha\beta\gamma\delta}R^{\alpha\beta\gamma\delta})^{-1/4}$. This identification first introduced in \cite{Moti & Shojai 2} and then studied in detail for Schwarzschild and Kerr spacetimes in \cite{Pawlowski}.
For FRW with $ R_{\alpha\beta\gamma\delta}R^{\alpha\beta\gamma\delta} = 12 (a^2\ddot{a}^2+(\dot{a}^2+Kc^2)^2)/(c^4 a^4) $ we would have  
\begin{align}
G_0 \rightarrow G(k) = G(\xi R_{\alpha\beta\gamma\delta}R^{\alpha\beta\gamma\delta}) & =\frac{G_{0}}{1+ \omega G_{0}\xi^2 (R_{\alpha\beta\gamma\delta}R^{\alpha\beta\gamma\delta})^{1/2}} \nonumber\\
& = \frac{G_{0}}{1+ 2\sqrt{3}\mathcal{ٰA}(a^2\ddot{a}^2+(\dot{a}^2+Kc^2)^2)^{1/2}/a^2 } \ ,
\end{align}
and the improved Friedmann equations
\begin{align}
  & 3\frac{\ddot{a}}{a}(1+ 2\sqrt{3}\mathcal{ٰA}\frac{(a^2\ddot{a}^2+(\dot{a}^2+Kc^2)^2)^{1/2}}{a^2}) =-4\pi G_{0}\rho( 1 + 3 w) \ , \\
  & (\frac{\ddot{a}}{a} + 2\frac{\dot{a}^2}{a^2} +2\frac{c^2 K}{a^2})(1+ 2\sqrt{3}\mathcal{ٰA}\frac{(a^2\ddot{a}^2+(\dot{a}^2+Kc^2)^2)^{1/2}}{a^2}) = 4\pi G_{0}\rho( 1 - w) \ .
\end{align}
The scale factor for flat universe filled with cosmological constant is $ a_4(t) = a_0 e^{\alpha_4(\tau)} $, with
\begin{equation}
 \alpha_4''( 1 + \mathcal{A}\frac{\sqrt{12}}{t_0^2}(\alpha_4''^2+2\alpha_4''\alpha_4'^2+2\alpha_4'^4)^{1/2})= 0 \ .
\end{equation}
Since $\mathcal{A}>0 $ the term in the parentheses does not vanishes, the solution is the classical de Sitter universe with  constant density
\begin{equation}
\rho_4 = \rho_i (1+2\sqrt{6} \mathcal{B}) \ .
\end{equation}
\end{itemize}

\begin{figure}[h]
        \centering
        \includegraphics[width=0.6 \textwidth]{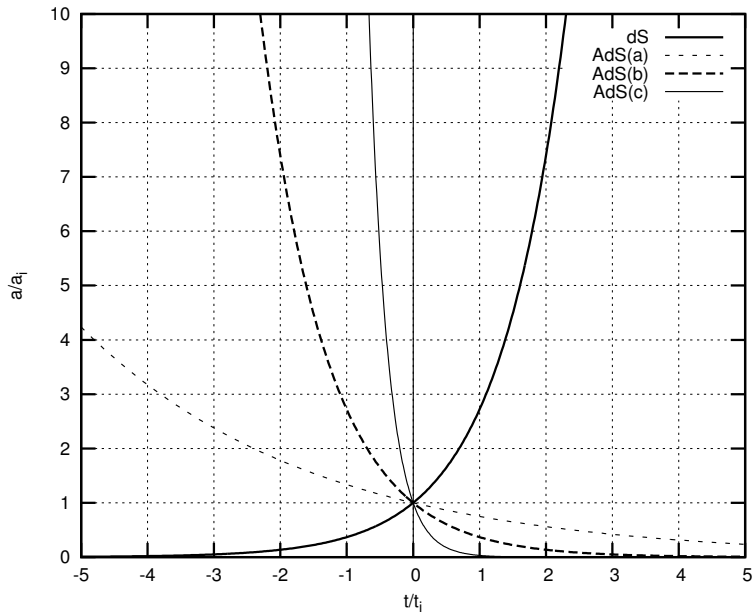}
        \caption{The bold solid line is the evolution of normalized de Sitter scale factor for all four cutoff identifications. The other three plots are the anti--de Sitter solution  obtained from cutoff identification with curvature scalar, for various values of $\mathcal{A}$ which depends on the quantization parameter $\omega$. We assumed $t_i=2\sqrt{3\mathcal{A}}$. Case (a) is drawn for $\mathcal{A}=4t_i^2$, (b) for $\mathcal{A}=t_i^2/3$ and (c) for $\mathcal{A}=t_i^2/36$.} 
        \label{IRWa-eps}
\end{figure}

In order to summarize the results, let us investigate the behaviour of scale factor and density for different cases. The plot of normalised scale factor as a function of normalised time is shown in figure \ref{IRWa-eps}. For all four cases, there is a de Sitter solution. There is also an anti--de Sitter solution, $a^{(\mathcal{A})}$, which is plotted for three different values of $\mathcal{A}$. 

\begin{figure} [h]
        \centering
        \includegraphics[width=0.6 \textwidth]{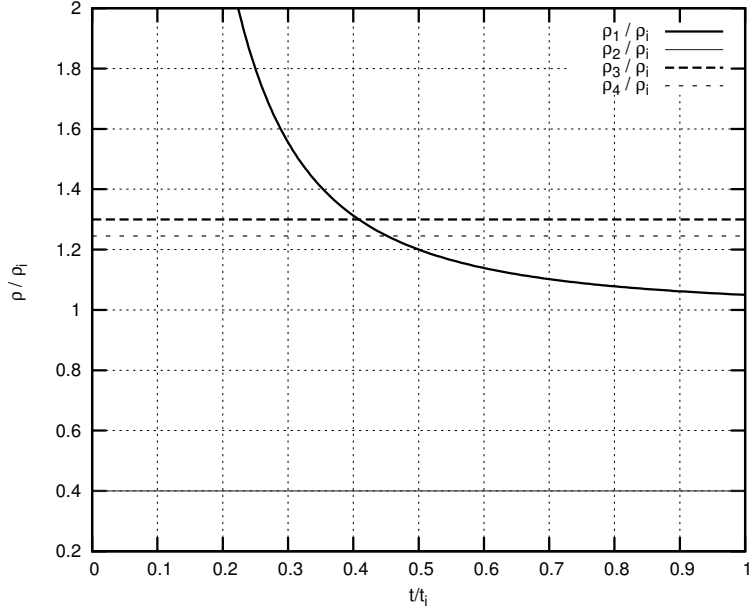}
        \caption{ The evolution of normalized densities $\rho_1/\rho_i$, $\rho_2/\rho_i$, $\rho_3/\rho_i$ and $\rho_4/\rho_i$. Contrary to the three cases of the mentioned cutoff identification which give constant densities, the evolution of  $\rho_1/\rho_i$ for standard identification  is time dependent and decreasing. Since the identification by squared scalar curvature constraints $\mathcal{B}$ to $\mathcal{B}<1/12$, all the curves are plotted for $\mathcal{B}=1/20$. }
        \label{IRWb-eps}
\end{figure}

In figure \ref{IRWb-eps} the evolution of improved matter densities is shown. For all three mentioned cases of identification we have constant density. In contrast, the matter density for naive identification decreases exponentially with time. 

\subsection{Parameter improvement}
This is the improving $G_0$ in the action to $G(x)$, but considering $G(x)$ as a background input field. The evolution of this background field is determined by ERGE. This evolution restricted the field equations obtained by extremizing the action \cite{Reuter & Weyer 1}. Although it is called action improvement in some papers, but the suitable name is parameter improvement, since it considers the couplings just as a parameter in the action. Clearly the general covariance breaks in this approach. To save the general covariance, some dynamical terms for this input field should be added to the action. Writing the appropriate dynamical terms is a hard and maybe impossible task, because they have to be such that the obtained equation of motion gives the same form for $G$ as the form given by ERGR.

As a result such an improvement method leads to an artificial model.

\subsection{Action improvement}
It is now clear that the best way of inclusion of quantum effects coming from the asymptotically safe gravity is to use the mentioned cutoff identification and improve the action. Since in this method of identification the couplings are functions of gravitational fields ($\chi_1\cdots\chi_{20}$), their dynamics are included in the resulting equations of motion, when extremizing the action.

For a typical running coupling where the cutoff momentum identified by a function of some curvature invariant, $\chi$, the improved Einstein--Hilbert action is
\begin{equation} 
  S_{IEH} = \frac{1}{16 \pi} \int \dd[4]{x} \frac{\sqrt{-g}}{G(\chi)}R \ .
\end{equation}
and the field equations come from the least action principle
\begin{equation}
\var{S} = \var{S_{IEH}}+\var{S_M} = 0
\end{equation}
in which $S_M$ is the matter action. 

In what follows, we shall derive the improved Einstein equation for the three cases $\chi=R^{-1/2}$, $\chi=(R_{\alpha\beta}R^{\alpha\beta})^{-1/4}$ and $\chi=(R_{\alpha\beta\gamma\delta}R^{\alpha\beta\gamma\delta})^{-1/4}$.
Although  $\chi$ may be a function of all twenty curvature invariants, but for simplicity and comparison of the results with other improvement methods, these three cases are enough.

\begin{itemize}

\item $\mathbf{\chi \equiv \chi_1 = (R)^{-1/2} :}$
The variation of the improved Einstein--Hilbert action would be
\begin{multline}
\var{S_{IEH}} = \frac{1}{16\pi} \int\dd[4]{x} \sqrt{-g} 
\Bigl( \bigl((R_{\alpha\beta}-\frac{1}{2}R g_{\alpha\beta})\mathcal{J}_1 -\frac{1}{2}\mathcal{K}_1RR_{\alpha\beta}\bigr)\var{g^{\alpha\beta}} + A^{\alpha\beta} \var{R_{\alpha\beta}} \Bigr)
\end{multline}
where $A^{\alpha\beta} = \bigl(\mathcal{J}_1 -\frac{R}{2}\mathcal{K}_1\bigr) g^{\alpha\beta} $, $\mathcal{J}_1 \equiv G^{-1}(\chi_1) $ and $ \mathcal{K}_1 \equiv  \frac{2\pdv*{G(\chi_1)}{\chi}}{G(\chi_1)^2} $.
Hence, considering the matter term, the field equation becomes
\begin{equation} \label{eq4.1}
G_{\alpha\beta} =  \frac{8\pi}{\mathcal{J}_1} T_{\alpha\beta} + \frac{1}{\mathcal{J}_1}X_{(1)\alpha\beta}
\end{equation}
where $ T_{\alpha\beta} \equiv  \frac{-2}{\sqrt{-g}} \frac{\var{S_M}}{\var{g^{\alpha\beta}}} $ is the energy--momentum tensor and $X_{(1)\alpha\beta}= \mathcal{K}_1 R R_{\alpha\beta}/2 + (\nabla_{\alpha}\nabla_{\beta} - g_{\alpha\beta}\square) \Bigl( \mathcal{J}_1-\mathcal{K}_1R/2\Bigr)$ contains the effects of improvement on the dynamics.

\item $\mathbf{\chi \equiv \chi_2 = (R_{\alpha\beta}R^{\alpha\beta})^{-1/4}:}$
For this identification, the variation of the improved Einstein--Hilbert action results in
\begin{multline}
 \var{S_{IEH}} =  \frac{1}{16\pi} \int \dd[4]{x} \sqrt{-g}
 \Bigl( ( R_{\alpha\beta} -\frac{1}{2}R g_{\alpha\beta} )\mathcal{J}_2  -\mathcal{K}_2 R R_{\lambda\alpha} R_{\gamma\beta}g^{\lambda\gamma} \\
- \nabla_{\rho} \nabla_{\beta} B_{\alpha}^{\rho} + \frac{1}{2} g_{\alpha\beta}\nabla_{\mu}\nabla_{\nu}B^{\mu\nu} + \frac{1}{2}\square B_{\alpha\beta} \Bigr) \var{g^{\alpha\beta}}
\end{multline}
where $\mathcal{J}_2 \equiv  G(\chi_2)^{-1} $, $ \mathcal{K}_2 \equiv  \frac{2 \pdv*{G(\chi_2)}{\chi_2}}{G(\chi_2)^2}$ and $ B^{\alpha\beta} = \mathcal{J}_2g^{\alpha\beta}- \mathcal{K}_2 R R^{\alpha\beta} $.
The improved equations of motion are thus
\begin{equation} \label{eq4.2}
  G_{\alpha\beta} = \frac{8\pi}{\mathcal{J}_2} T_{\alpha\beta} + \frac{1}{\mathcal{J}_2}X_{(2)\alpha\beta}
\end{equation}
where
\begin{multline}
X_{(2)\alpha\beta} = 
\bigl(\nabla_{\alpha}\nabla_{\beta} - g_{\alpha\beta} \square\bigr) \mathcal{J}_2
+ \mathcal{K}_2 R R_{\alpha\mu} R_{\beta\nu} g^{\mu\nu} \\   
- \nabla^{\sigma}\nabla_{\beta}( \mathcal{K}_2 R R_{\alpha\sigma})
 +\frac{1}{2}\square( \mathcal{K}_2 R R_{\alpha\beta}) + \frac{1}{2} g_{\alpha\beta} \nabla_{\rho}\nabla_{\sigma} ( \mathcal{K}_2 R R^{\rho\sigma}) \ .
\end{multline}

The black hole solution of this equation is studied in \cite{Moti & Shojai 2}. It is shown that the thermodynamics of such a black hole leads to  different results.

\item $\mathbf{\chi \equiv \chi_3 = (R_{\alpha\beta\gamma\delta}R^{\alpha\beta\gamma\delta})^{-1/4}:}$ 
For this case we have 
\begin{multline} \label{eq4.4}
\var{S} =  \frac{1}{16\pi}\int \dd[4]{x} \Bigl( \var{(\sqrt{-g})}\frac{R}{G(\chi_3)}
   + \frac{\sqrt{-g}}{G(\chi_3)} \var{R} 
    -\sqrt{-g} R \frac{\var{\chi_3}}{G(\chi_3)^2} \pdv{G(\chi_3)}{\chi_3} \Bigr) + \var{S_M} \ .
\end{multline}
On using the relation
\begin{multline}
 \var{(R_{\alpha\beta\gamma\delta}R^{\alpha\beta\gamma\delta})} =  2 R_{\alpha\beta\gamma\delta}R^{\ \beta\gamma\delta}_{a} \var{g^{\alpha a}} + 2 R^{\alpha\beta\gamma\delta} \nabla_{\delta}\nabla_{\alpha}\var{g_{\gamma\beta}} \\ + 2R^{\alpha\beta\gamma\delta} \nabla_{\gamma}\nabla_{\delta}\var{g_{\beta\alpha}} + 2 R^{\alpha\beta\gamma\delta} \nabla_{\gamma}\nabla_{\beta}\var{g_{\delta\alpha}}
\end{multline}
in the equation \eqref{eq4.4}, we get
\begin{equation} \label{eq4.3}
G_{\alpha\beta} = \frac{8\pi}{\mathcal{J}_3} T_{\alpha\beta} + \frac{1}{\mathcal{J}_3}X_{(3)\alpha\beta}
\end{equation}
where
\begin{multline}
X_{(3)\alpha\beta} = (\nabla_{\alpha}\nabla_{\beta} - g_{\alpha\beta}\square)\mathcal{J}_3 + \mathcal{K}_3 R_{\alpha\gamma\delta\lambda}R^{\ \gamma\delta\lambda}_{\beta} - \\
g_{\alpha c}g_{\beta b}(\nabla_a\nabla_d-\nabla_d\nabla_a)R^{abcd}R\mathcal{K}_3 - g_{\alpha b}g_{\beta a} \nabla_d \nabla_c (\mathcal{K}_3 RR^{abcd}) 
\end{multline}
is the correction to the dynamics of gravitational field.
\end{itemize}

Clearly all the cases can be written as $G_{\alpha\beta} = 8\pi T_{\alpha\beta}/\mathcal{J}_i + X_{(i)\alpha\beta}/\mathcal{J}_i$, where $i=1,2 \ \text{or} \ 3$.
The factor $1/\mathcal{J}_i$ besides $T_{\alpha\beta}$ includes direct influence of the running couplings on the equations of motion, while the term $X_{(i)\alpha\beta}$ is the corrections to the dynamics. 
 
\section{Conclusion} \label{4}
The Weinberg's \textit{asymptotic safety conjecture} is able to save the theory from divergences, if its running coupling constants tend to a non--Gaussian fixed point at the UV limit. Exact renormalization group method is a successful method in search for such a non--Gaussian fixed point. Using this method, the effective average action is truncated up to the desired terms.
The initial condition for this equation is the classical action with running coupling constants.
Hence, one can find the running coupling constants using the ERGE.
This method predicts anti--screening terms for the running gravitational coupling constant.

Quantum improvement of the classical solutions needs some kind of cutoff identification with some space--time quantity. It is common to use the inverse of some physical length for cutoff identification, but the notion of length needs some attention for curved space--times. We propose here to use tidal forces and thus the Riemann curvature tensor as the measure of smallness of distances to have a proper cutoff identification. Hence, improving the running coupling constant in the action in a dynamical way gets possible \cite{Bonanno, Moti & Shojai 2}.

As mentioned, the process of improving the gravitational coupling constant to the running one can be classified in four cases: solution improvement, equation of motion improvement, parameter improvement and action improvement.

Although the backreaction of the quantum corrections could enter the improved solution to some extent, using an iterative method \cite{iterative}, but since they have no counterpart in the action, considering them as dynamical effects is a debatable issue. Hence, the solution improvement method could not be recommended unless the backreaction effects of the quantum corrections on the background geometry is negligible. Indeed, the parameter improvement is accompanied by the breakdown of general covariance unless introducing the related dynamical term.

By the equation of motion improvement, the improvement is done at the level of equations of motion obtained from the classical action. Although the evolution of running couplings is fixed by the ERGE, the dynamical effects of the quantum corrections on the gravitational field are ignored in this method.

On the other hand, in the action improvement method besides saving the general covariance, the dynamics of these corrections would appear in the obtained equation from the improved action. Hence, the action improvement seems to be the most suitable improvement method. But, complexity of improved equation would be a serious challenge in this method. Indeed, the correction terms which appear in the equation of motion would depend on the chosen identification function which has length inverse dimensionality. If we consider the corrections to be small enough, using the classical solution as the first step of iteration, would be sufficient for finding the proper corrected solution. This method would confine the  chosen curvature invariants for identification to the ones which have non--singular value at the classical limit.

Here the results of various cutoff identifications for equation of motion and action improvements are studied for a simple cosmological model. In addition to the classical solution, we observed that pure quantum solutions are also possible.

Also, the field equation of the improved action via various cutoff identifications is derived. The general improved equation of motion is $G_{\alpha\beta} = 8\pi T_{\alpha\beta}/\mathcal{J}_{i} + X_{(i)\alpha\beta}/\mathcal{J}_{i}$, where the index $i$ discriminates between different cutoff identifications. The term containing $ X_{(i)\alpha\beta}$ reflects the quantum effects that are even present for vacuum solutions. Since $X_{(i)\alpha\beta}$ is a function of $\mathcal{K}_i$ and its derivatives and the derivatives of $\mathcal{J}_i$, it can be interpreted as a dynamical effect of quantization which are carried out by the improved $G(\chi)$.

At this end, it should be noted that the cutoff identification is a function of the quantum corrected metric, which is a solution of the improved equation of motion. Also the equation of motion cannot be determined uniquely until the improvement method is applied. Therefore, it can be deduced that the improvement process and the cutoff identification are correlated because of the additional dynamical terms.

\end{document}